\documentclass[pra,preprint,showpacs,preprintnumbers,amsmath,amssymb]{revtex4-1}

\usepackage{graphicx}
\usepackage{dcolumn}
\usepackage{bm}

\begin{document}

\preprint{Version 02}

\title{Maxwell-Bloch modeling of an x-ray pulse amplification in a one-dimensional photonic crystal}

\author{O. Peyrusse}
 \email{olivier.peyrusse@univ-amu.fr}
\affiliation{Aix-Marseille Univ., CNRS UMR 7345, PIIM Marseille, France}

\author{P. Jonnard, J.-M. Andr{\'e}}
\affiliation{Sorbonne Universit{\'e}, Facult{\'e} des Sciences et Ing{\'e}nierie, UMR CNRS,
Laboratoire de Chimie Physique -
Mati{\`e}re et Rayonnemment, 4 Place Jussieu, F-75252 Paris Cedex 05, France}

\date{\today}

\begin{abstract}
We present an implementation of the Maxwell-Bloch (MB) formalism for the study of x-ray emission
dynamics from periodic multilayer materials whether they are artificial or natural.
The treatment is based on a direct Finite-Difference-Time-Domain (FDTD) solution of Maxwell
equations combined with Bloch equations incorporating a random spontaneous emission noise. Besides
periodicity of the material, the treatment distinguishes between two kinds of layers, those being
active (or resonant) and those being off-resonance. The numerical model is applied to the problem
of $K\alpha$ emission in multilayer materials where the population inversion could be created by fast
inner-shell photoionization by an x-ray free-electron-laser (XFEL). Specificities of the resulting
amplified fluorescence in conditions of Bragg diffraction is illustrated by numerical simulations.
The corresponding pulses could be used for specific investigations of non-linear interaction of
x-rays with matter.

\end{abstract}


\maketitle
\section{Introduction}

Non-linear optical devices (NLO) have been a vivid subject of study for their numerous applications.
Within the domain of x-ray quantum optics \cite{Adams13,Adams},
the field of non-linear x-ray (NLX-ray) devices is much less explored since, compared with the optical
range, the control of x-rays is more difficult. The simplest NLX-ray device is an ensemble of 2,3-level
atoms for which different studies of pulse propagation and of several non-linear effects have been
reported (see for instance \cite{Sun10}). Another typical NLX-ray devices are multilayer materials which are used in
x-ray optics. Short and ultra-intense x-ray sources such as x-ray free electron lasers
(XFELs) are pushing the boundaries of the
response to x-rays in such devices. Besides this,
it has been proved that XFEL sources have the potential to create
large population inversions in gases \cite{Roh12}, clusters \cite{Ben20},
solids \cite{Beye13,Yon15,Jon17} and liquids \cite{Kroll18}, resulting in the creation
of an x-ray amplifying medium.
These approaches based on lasing in atomic media have an important potential
to obtain useful short and coherent x-ray pulses. Indeed, high-quality short pulses going beyond
the inherent defaults of SASE XFEL pulses (of spiky and chaotic nature)
are a prequisite for future investigations concerning x-ray quantum optics, x-ray scattering,
precision spectroscopy or pump-probe experiments requiring a coherent probe.
Compared with conventional lasers, these approaches suffer from the lack of a resonator to extract most of
the energy stored in an inverted medium.
In other words, there remains the problem of realizing
x-ray feedback to achieve laser oscillation in the x-ray range.
Hence a work going in that direction has recently been reported \cite{Halavanau20}.
In this reported work a classical multipass meter-sized laser cavity has
been set up. The x-ray lasing medium being a liquid jet pumped by an XFEL.
Besides this, within the context of XFEL excitation and to extract energy stored in an inverted medium,
the idea of using the phenomenon of collective spontaneous decay or superradiance
(named also superfluorescence) has been discussed and explored \cite{Naga11} but in the visible range.
Independently, it has been suggested that a laser action in the x-ray range
can be provided by Bragg reflection inside a natural crystal or inside an artificial multilayer material
\cite{Yariv77,JMA14,Pey20}. Note that in the first case, Bragg condition is constrained by the crystal periodicity.

The goal of this paper is to study numerically x-ray feedback under Bragg conditions
as well as pulse propagation in 1D photonic crystals in which a population inversion has been initiated
by some external source.
Here we go beyond a description where the multilayer is simply described by the complex
refractive index of each layer \cite{Pey20}. Even when the complex part of the refractive index is
negative (i.e. amplifying), such a description is basically linear and corresponds to the linear phase of the
interaction of an x-ray pulse with the material. Note that this remark concerns the
{\it active} layers only. {\it Passive} layers, for which no resonant response is
expected, can still be described by a complex refractive index. This defines the specificity
of our description which mixes a non-linear treatment and a linear treatment, i.e. more precisely
using a Maxwell-Bloch (MB)
formalism or a standard formalism,
depending on the kind of layer (active or passive).

In this paper we consider a large number of photons in the radiation modes. As a
consequence, quantum fluctuations are neglected and the electromagnetic (EM) field is described
by Mawxell equations. In the absence of an external source, one shortcoming of the MB description is
that there is no mechanism  for spontaneous emission. It is well-known that this problem can
be overcome by adding a phenomenological fluctuating polarization source that simulates
spontaneous emission (although this approach has some drawbacks as discussed below).
Compared with many calculations of x-ray lasing in gas or plasmas
(see for instance Refs \cite{MGill76, Lar00, Wen14, Lyu20})
short spatial scales
involved in this multilayer context, do not permit the use of the {\it slowly varying envelope
approximation} so that basic Maxwell equations have to be solved directly. This is done here using
the so-called Finite-Difference-Time-Domain (FDTD) method \cite{Taflove}. Futhermore, in this multilayer
(or 1D photonic crystal) context, we consider a 1D plane geometry. Also, we consider only
two levels resonantly coupled to the EM field in the MB system. Other levels are taken into
 account through relaxation and source terms in the equations governing
populations of these two levels.

In the following, we present the physical model used here (Sec. II), underlining the specific
choices made for considering 1D photonic crystals in the x-ray range. Then, we
turn to a discussion of simulation results in Sec. III.
Physical situations considered here evolve gradually from very formal situations to
situations close to actual experimental conditions.
More precisely, we begin with a situation intended at testing the FDTD implementation in the
context of the fluorescence of a multilayer. Here there is no solving of the Bloch equations,
instead a source of emission is assumed in each cell (Sec. III-A).
Then,
after these considerations on the
validity of the FDTD implementation, we turn to MB calculations and
we consider in detail the problem of an x-ray pulse
propagation in a particular stack of bi-layers (Mg/Co) in which an initial population inversion is
supposed (Sec. III-B). After this, one considers the self-emission of such a stack, i.e. as initiated
by spontaneous emission (Sec. III-C). Finally, we turn to situations where an XFEL source is used
for pumping (i.e. for creating the population inversion) either a multilayer (such as a
bi-layer (Mg/Co) stack, Sec. III-D) or a simple Ni crystal (Sec. III-E).
Section IV summarizes these results.

\section{Theoretical approach}

\subsection{Basic equations}
As said in the introduction, the medium considered here consists of alternating {\it active} and {\it passive}
materials with a given periodicity.
{\it Active} in the sense of resonantly coupled with the EM field at some pulsation $\omega_o$
and {\it passive} if there is no resonant coupling.
The two first basic equations of our approach are the Faraday and the Ampere's laws, respectively written
in the form (SI units)

\begin{equation}
\vec\nabla \times \vec{E} = - \partial_{t} \vec{B}
\end{equation}

\begin{equation}
\frac{1}{\mu_{o}}\vec\nabla \times \vec{B} = \epsilon_{o} \epsilon_{r} \partial_{t} \vec{E} + \vec{j}
\end{equation}
$\vec{E}$, $\vec{B}$ are the electric and the magnetic fields (real quantities), respectively.
$\epsilon_{o}$, $\mu_{o}$ are the vacuum permittivity and permeability, respectively.
$\epsilon_{r}$ is the
relative permittivity (here time-independent)
and $\vec{j}$ is the local current induced by the EM field. In a linear
material, i.e. here in a {\it passive} layer, $\vec{j} = \sigma {\vec{E}}$ where $\sigma$ is
the electric conductivity. At pulsation $\omega_{o}$, adiabatic properties
of the material (in the sense of an instantaneous response to the applied field)
are included in the {\it real} quantities $\epsilon_{r}$ and $\sigma$. If the
material is described by a complex refractive index of the form (as in current data tables
\cite{Hen93, CXRO}),
\begin{displaymath}
n = (1-\delta)- i \beta,
\end{displaymath}
there is an equivalence between the conductivity approach and the refractive index in the sense that
\begin{align*}
& \epsilon_{r}=(1-\delta)^{2} - \beta^{2}\\
& \sigma= 2(1-\delta) \beta \omega_{o} \epsilon_{o}
\end{align*}
In a so-called {\it active} medium such as the so-called {\it active} layers, one has
$\vec{j}=\partial_{t} \mathcal{\vec{P}}$ where $\mathcal{\vec{P}}$ is the macroscopic polarization.
Considering two levels coupled by the EM field,
the macroscopic polarization is defined as the trace of the operator $N\rho  \vec{d}$ where $N$ is the
density of polarizable atoms, $\vec{d}$ is the atomic dipole and $\rho$ is the density matrix. Hence

\begin{equation}
\mathcal{\vec{P}}  = 2 N \text{Re}[\rho_{21}] \vec{d}
\end{equation}

\noindent The non-diagonal matrix element of the density matrix $\rho_{21}$ is a complex number as
the diagonal elements $\rho_{11}$, $\rho_{22}$ are real quantities. These matrix elements
are obtained
from the following evolution of the density matrix (Liouville equation)
$i \hbar \partial_t = [H, \rho] + \text{relaxation/source terms}$. Here, assuming a polarization along
axis Ox, the Hamiltonian $H$ has the form
$H=\begin{pmatrix}
\epsilon_{1} & -dE_{x}\\
-dE_{x} & \epsilon_{2} \\
\end{pmatrix}$
where both $E_{x}$ and $d=<1|d_{x}|2>$ are real quantities. $\epsilon_{1,2}$ are the energies of the coupled
levels. Hereafter, $\hbar\omega_{o}= \epsilon_{2} - \epsilon_{1}$. Level scheme of the problem is depicted
in Fig. 1.

\begin{figure}
\rotatebox{0}{\resizebox{7.0cm}{5cm}{\includegraphics{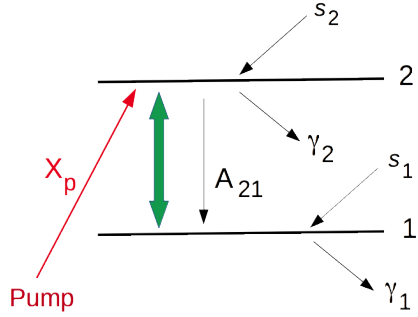}}}
\caption{\label{fig1} (Color online)
Level scheme with the associated source or loss rates. The thick line with 2 arrows
represents the Maxwell-Bloch coupling.
}
\end{figure}

\noindent In a warm or hot medium,
$\gamma_{1}$, $\gamma_{2}$ are the rates (in s$^{-1}$) of the collisional, radiative and Auger processes
which depopulate states, except
stimulated emission and spontaneous emission.
$A_{21}$ is the spontaneous emission rate. Here
$s_1$, $s_2$ are the possible population fluxes due to the all processes, except absorption between
states 1 and 2.
$\gamma_{1,2}$ and $s_{1,2}$ may involve other levels in the
system. We define
$\gamma_{\perp}=\frac{1}{2}(\gamma_{1} + \gamma_{2})+ \gamma_{\phi}$ so that $1 / \gamma_{\perp}$ can be seen
as the life time of the coherent superposition of states $|1>$ and $|2>$. In principle, $\gamma_{\phi}$ is
supposed to be the rate of events perturbing the wave-function without inducing a decay of eigenstates.
Hereafter, we called populations of states 1 and 2 the macroscopic quantities $N_{1}=N\rho_{11}$,
$N_{2}=N\rho_{22}$, respectively.

\noindent In an active layer, the set of equations to be solved locally for the populations
$N_{1}$, $N_{2}$ and for $P=N\rho_{21}$,
is then,
\begin{equation}
\partial_{t} N_{1} = - \frac{2}{\hbar} d E_{x} \text{Im}[P] + s_{1} - \gamma_{1} N_{1} + A_{21} N_{2}
\end{equation}

\begin{equation}
\partial_{t} N_{2} = \frac{2}{\hbar} d E_{x} \text{Im}[P] + s_{2} - \gamma_{2} N_{2} - A_{21} N_{2} + X_{p}
\end{equation}

\begin{equation}
\partial_{t} P = - i \omega_{o} P - \gamma_{\perp} P - i(N_{2}-N_{1}) dE_{x} + S
\end{equation}
where we added in Eq. (5), a pump source term $X_{p}$.
$S$ is a phenomenological random source modeling the spontaneous emission. In the absence of
external incoming radiation, $S$ acts as an energy seed for energy injection in the system.
Eqs.(1)-(6) correspond to our set of Maxwell-Bloch equations where Eqs. (4)-(6) concern the
active layers only.
In the literature, the name of Maxwell-Bloch \cite{Scully}, or
sometimes Maxwell-Schr\"{o}dinger \cite{MGill76}, is often given to the coupling of complex slowly varying
{\it envelopes} of the EM field with the Bloch equations for the density matrix.
Here, there is no approximation concerning the
field variation both in space and time.

\subsection{Wave equations in a 1D photonic medium at oblique incidence}

A sketch of the physical problem is given in Figure 2. The medium is made of a periodic stack of different
materials (at least two different materials). One is the active material (i.e. described by the MB
equations), the other(s) is(are) the passive (or linear) material(s), i.e. described by complex
refractive indices.

\begin{figure}
\rotatebox{0}{\resizebox{7.0cm}{5cm}{\includegraphics{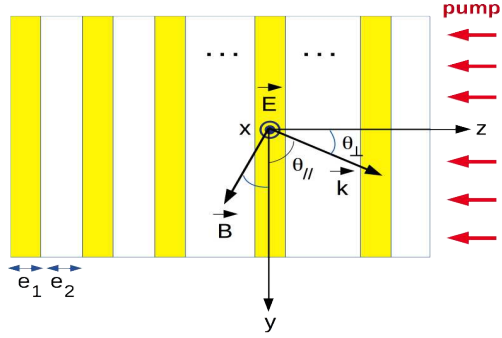}}}
\caption{\label{fig1} (Color online)
Sketch of a stack of bilayers made of one {\it active} element (yellow) and of one {\it passive} element.
On one side the stack is submitted to a source of excitation (the pump). Propagation of the emerging
radiation is described by Maxwell equations at some frequency (which is resonant for the active layers).
In active layers, response is described by the Bloch equations while in passive layers, response is given
by the refractive index.
}
\end{figure}

\noindent ($\vec{k}, \vec{E}, \vec{B}$) represent a plane-wave propagating in the material under the angle
$\theta_{\bot}$. Hereafter we make use of $\theta_{\bot}$ which is the angle with respect to a direction
{\it perpendicular} to surface $xy$ while it may be more convenient to use
$\theta_{//} = \frac{\pi}{2} - \theta_{\bot}$.
In this geometry, Faraday's law (Eq. (1)) reads,

\begin{align}
& \partial_{z} E_{x} = - \partial_{t} B_{y}\\
& \partial_{y} E_{x} = \partial_{t} B_{z}
\end{align}
while the Ampere's law (Eq. (2)) reads

\begin{equation}
- \partial_{z} B_{y} + \partial_{y} B_{z} = \frac{\epsilon_{r}}{c^{2}} \partial_{t} E_{x} + \mu_{o} j_{x}
\end{equation}
where $j_{x}=\sigma E_{x}$ in a {\it passive} layer. In an {\it active} layer, $\epsilon_{r} =1$ and
$j_{x}=\partial_{t} \mathcal{P}_{x}$, $\mathcal{P}_{x}$ being deduced from Eq. (3), i.e.
$\mathcal{P}_{x} = 2d\text{Re}[P]$.
Translating (7), (8), (9) from the time-domain to the frequency-domain gives three equations in
which $E_x$ has the behavior $E_{x} \sim exp(i\omega_{o}t \pm i k y \sin{\theta_{\bot}})$. Eliminating $B_{z}$ and
going back to the time domain ($i\omega_{o} \rightarrow \partial_{t}$)
finally gives the following two equations governing the behavior of a plane-wave
for an arbitrary oblique incidence in the multilayer material,

\begin{align}
& \partial_{z} E_{x} = - \partial_{t} B_{y}\\
& - \partial_{z} B_{y} = \frac{1}{c^2} (\epsilon_{r}-\sin{^{2}\theta_{\bot}})\partial_{t} E_{x} + \mu_{o}j_{x}.
\end{align}

\noindent To solve these equations, one uses the usual FDTD method \cite{Taflove}
namely a second-order central
difference scheme introduced by Yee \cite{Yee66}.
Yee's scheme consists in writing the central differences of $E_{x}$ and $B_{y}$,
shifted in space by half a cell and in time by half a time step. In our implementation, $B$ is evaluated
at the edge of each cell while $E$ is evaluated at the center. Accordingly, proper boundary conditions for $B$
have to be applied on each side of the multilayer. Considering that the two external layers (on both sides)
correspond to the vacuum (refractive index $n=1$), and in order to remove reflections from these boundaries, we used
second-order Absorbing Boundary Conditions (ABC) \cite{ABC}. Together with $E_{x}$, the quantities
$N_{1}$, $N_{2}$ and $P$ are cell-centered and are advanced in time using a Crank-Nicolson scheme.\\
Concerning the sampling in space, typically at least 20 steps per wavelength are necessary. This involves
a subdivision of each layer in much smaller layers of thicknesses $\Delta z$. Accordingly, the sampling in
time is governed by the Courant limit. More precisely, an inspection of Eq. (11) shows that the front
phase velocity is $c/\sqrt{\epsilon_{r}-\sin^{2}{\theta_{\bot}}}$. Then the time-step must be such that

\begin{displaymath}
\Delta t \le \frac{\Delta z}{c} \sqrt{\epsilon_{r}-\sin^{2}{\theta_{\bot}}}.
\end{displaymath}

\subsection{Incident source - Spontaneous emission}

For some applications, it may be useful to consider the seeding of the multilayer by some
incident external polarized X-ray pulse (see Sec. III-B below).
Practically, this can be an external source of X-rays at $\omega_{o}$
generated independently.
Note that we distinguish this potential seeding source at $\omega_{o}$ from another source
(not at $\omega_{o}$) allowing to create a population inversion.
In order to follow an X-ray pulse propagating in one specific direction,
the positive z-direction for instance (see Fig. 2),
this source at $\omega_{o}$ has to be properly implemented. As
usual in FDTD simulations, this is accomplished using a total-field/scattered-field (TFSF) boundary
\cite{Merewether80} at
the point where the source(s) is(are) put. For instance, this source can be placed in the left vacuum cell of
our simulation domain. More precisely, one defines both an incident electric field $E^{inc}$ and an
incident magnetic field $B^{inc} = \cos{\theta_{\bot}}\sqrt{\epsilon_{r}} E^{inc}$
($\epsilon_{r}=1$ in the vacuum).
At the location of the source, according
to the TFSF method, $E$ must be replaced by $E-E^{inc}$ in the discretized evolution of the B field while
$B$ must be replaced by $B+B^{inc}$ in the discretized evolution of the E field.\\

Independently of any external source at $\omega_{o}$, the source of spontaneous emission in Eq. (6)
(the term $S$)
can be modeled as a
Gaussian white noise following the guidelines of Ref. \cite{Lar00}, an approach followed later by others
\cite{Wen14, Lyu20}. The interest of this approach is that it
provides the correct spectral behavior for the field \cite{Lar00}.
Here, one starts from the simplified (local) system (see Eq. (11) and Eq. (6)) coupling the electric field
and $P=N\rho_{21}$

\begin{align}
&\frac{dE}{dt} = \alpha \text{Re}[\frac{dP}{dt}] \ \ \ \ \ \ \ \ \ \ \ (\alpha=-2\mu_{o} c^{2} d)\\
&\frac{dP}{dt} = - i\omega_{o} P - \gamma P + S
\end{align}
where the noise source $S$ (complex) has the correlation function $\langle S^{*}(t') S(t) \rangle = F \delta(t'-t)$.
The notation $\langle...\rangle$ is used to represent the statistical ensemble averaging and $F$ is a constant defined
by the following arguments. From the density of the electric field $\frac{\epsilon_{o}}{2} E^{2}$, one
defines an average power density  $W$ which must be equal to the power emitted by spontaneous emission (in
one direction) so that,

\begin{equation}
W=\frac{d}{dt} (\frac{\epsilon_{o}}{2} \langle E^{2} \rangle) = \frac{1}{4\pi}N_{2} A_{21} \hbar\omega_{o}
\end{equation}

\noindent From the formal solution of Eq. (12),
$E(t)= \alpha \int_{-\infty}^{t} \text{Re}[\frac{dP(t')}{dt'}] dt'$, one gets\\

$W=\epsilon_{o} \langle E\frac{dE}{dt} \rangle = \alpha^{2}\epsilon_{o} \int_{-\infty}^{t}
\langle \text{Re}[\frac{dP(t')}{dt'}] \text{Re}[\frac{dP(t)}{dt}] \rangle dt'$. Using (13), $W$ becomes,

\begin{equation}
W=\frac{\epsilon_{o}}{4} \alpha^{2} \int_{-\infty}^{t}
\left ( \langle  \frac{dP^{*}(t')}{dt'} \frac{dP(t)}{dt} \rangle
+\langle  \frac{dP(t')}{dt'} \frac{dP^{*}(t)}{dt} \rangle \right ) dt'.
\end{equation}
From the formal solution of Eq. (13),
$P(t)=\int_{-\infty}^{t} S(t')e^{-i\omega_{o}(t-t')}e^{-\gamma(t-t')} dt'$, it is easy to
calculate quantities $\langle...\rangle$, so that after averaging over one period, calculation of
the integrals in (15) gives (since $\omega_{o} >> \gamma$)
$W=\frac{\epsilon_{o}}{4} \alpha^{2} \frac{F}{2} \frac{\omega_{o}^{2}}{\gamma^{2}}$. Then from relation (14), one
gets

\begin{equation}
F(z,t)=\frac{2 A_{21} \hbar \omega_{o} N_{2}(z,t)}{\alpha^{2} \pi \epsilon_{o}} \frac{\gamma^2}{\omega_{o}^2}.
\end{equation}

\noindent Practically, over a time step $\Delta t$, the noise source term $S$ in Eq. (6) is a random
complex number $u +i t$ distributed according to the law
$\frac{1}{\pi\sigma_{S}^2} \exp{ -(u^{2} + t^{2})/\sigma_{S}^{2})}$ with $\sigma_{S}=\sqrt{F\Delta t}$.

\section{Simulations results}
A numerical code based on the model described above has been built. The {\it active} materials considered in
this article are K-shell photoionized
magnesium (Secs III-A,B,C,D) or nickel (Sec. III-E).
In magnesium, according to the level scheme depicted in Fig. 1, level 2
stands for $1s\ 2s^{2}\ 2p^{6}\ [3s^{2}]$ and level 1
stands for $1s^{2}\ 2s^{2}\ 2p^{5}\ [3s^{2}]$. In nickel, level 2
stands for $1s\ 2s^{2}\ 2p_{1/2}^{2}2p_{3/2}^{4}\ 3s^{2} 3p^{6} 3d^{8}\ [4s^{2}]$ and level 1
stands for $1s^{2}\ 2s^{2}\ 2p_{1/2}^{2}2p_{3/2}^{3}\ 3s^{2} 3p^{6} 3d^{8}\ [4s^{2}]$.
Compared with neutral atoms, outer electrons
in solid Mg or Ni (denoted by
[...]) are delocalized.
In what follows, we either set populations 2 and 1 (likewise the density of inversion) (Secs III-B,C)
or we explicitely
consider a time-dependent pumping (Secs III-D,E).

\noindent In this last case,
initial atoms (in the state $|0\rangle \equiv 1s^{2}\ 2s^{2}\ 2p^{6}\ [3s^{2}]$ for Mg or
$[Ar] 3d^{8}\ [4s^{2}]$ for Ni)
are photoionized by an external X-ray source (supposedly an XFEL beam) hereafter named as "the pump",
tuned above
the {\it K} edge. This pumping results in the population of the core-excited state $|2\rangle$ radiatively
coupled to state $|1\rangle$ by the decay $2p \rightarrow 1s$. In conditions of weak pumping, this
coupling corresponds to the usual $K \alpha$ fluorescence. Note that in conditions of weak pumping,
state $|2\rangle$ predominantly decays via
Auger decay (with the rate $\Gamma_{2}$)
while both states $|1,2\rangle$ are also affected by the photoionizing pump. In Mg, we neglect
the fine-structure splitting of the $K \alpha$ line since it is smaller than the Auger width.
For Ni, the splitting $K \alpha_{1} - K \alpha_{2}$ exceeds largely the Auger width and we chose to
consider the $K \alpha_{1}$ line only.
Levels $|1,2\rangle$ are the two levels considered in our Maxwell-Bloch modeling. According to
Fig. 1 and Fig. 2, quantities $\gamma_{1}$, $\gamma_{2}$ and $X_{p}$ depend on the local intensity
of the pump $I_{p}$ in the sense that

\begin{align}
& \gamma_{1}(z,t)= \sigma_{1s} \ \frac{I_{p}(z,t)}{h\nu_{p}}\\
& \gamma_{2}(z,t) = \Gamma_{2}\\
& X_{p}(z,t) = \sigma_{1s} N_{o}(z,t)\ \frac{I_{p}(z,t)}{h\nu_{p}}
\end{align}
where $\sigma_{1s}$ denotes the $1s$ photoionization cross-section at energy $h\nu_{p}$.
$I_{p}$, $h\nu_{p}$ are the
intensity (here a power per surface unit)
and the photon energy of the pump, respectively. $N_{o}$ is the population density of
state $|0\rangle$. If $h\nu_{p}$ is greater than the second 1s ionization threshold, the term
$\frac{\sigma_{1s}}{2} \ \frac{I_{p}(z,t)}{h\nu_{p}}$ must be added to the
right side of Eq. (18).
At this step it is important to remark that, for a normal incidence and being off an accidental
situation where the period of the material would be an integer of $\lambda/2$ (i.e. off-Bragg),
one may adopt
for the pump the simple {\it corpuscular} point of view of photon absorption. Hence,
for a pump propagating from the right (see Fig. 2),
the pump intensity obeys the following photon transport equation

\begin{equation}
\frac{1}{c}\frac{\partial I_{p}(z,t)}{\partial t} - \frac{\partial I_{p}(z,t)}{\partial z} =
- k_{p} (z,t) \ I_{p}(z,t)
\end{equation}
with

\begin{align*}
k_{p} (z,t) &= \sigma_{1s} N_{o}(z,t) + \sigma_{1s} N_{1}(z,t)
+ \frac{\sigma_{1s}}{2} N_{2}(z,t)\ \ \  \text{(in the active material)}\\
&= \sigma^{passive} N^{passive}(z,t) \ \ \ \ \ \  \ \ \ \ \ \  \text{(in the passive material)}
\end{align*}
in which $N^{passive}(z,t)$ is the atom density in a {\it passive} layer, $\sigma^{passive}$
being the corresponding absorption cross section at $h\nu_{p}$.
Finally, the population of state $|0\rangle$ evolves as

\begin{equation}
\frac{\partial N_{o}(z,t)}{\partial t} = - \sigma_{1s} N_{o} (z,t) \frac{I_{p}(z,t)}{h\nu_{p}}.
\end{equation}

\noindent In the case where the pump is explicitely taken into account,
equations (17)-(21) have to be solved simultaneously with the previous Maxwell-Bloch set of
equations.

In the following, after checking the right behavior of the FDTD implementation (Sec. III-A),
we describe the propagation of an X-ray pulse at the $K \alpha$ energy for
different situations of increasing complexity whether the pulse is of external origin (Sec. III-B)
or not
(i.e. originating from spontaneous emission, Secs III-C,D,E). Note that, in Secs III-D,E we
explicitely consider the pumping by an external photoionizing x-ray source.

\subsection{Simple propagation in a multilayer material}
A first and minimal implementation amounts to considering that all the layers are of {\it passive} nature, i.e.
simply described by a complex refractive index. The goal is to assess the necessary number of subdivisions of each
layer in our specificic problem of wave propagation in the x-ray range, in a stratified medium made of nm-size layers.
Indeed, this number of subdivisions
defines a typical space interval $\Delta z$ on which Maxwell equations are discretized according
to the FDTD scheme mentioned above.
Of course, this defines the time step $\Delta t$ as discussed in Sec. II-B. As a test case,
we consider here a sample already considered in a context of synchrotron irradiation \cite{Jon14}.
It consists in a stack of 30 bilayers (Mg/Co) of thicknesses $e_{1} = 5.45 \ \text{nm}$ and $e_{2} = 2.55 \ \text{nm}$,
respectively so that the whole stack is denoted (Mg/Co)$_{30}$.
In each Mg layers, a source of radiation is supposed to emit at the $K \alpha$ line energy (1253.6 eV). These
sources are all the same and of the form
$E(t)=A \exp{[-0.5 (t-t_{o})^{2}}/\tau_{p}^{2}] \sin{\omega_{o}t}$ (where $A$ is an arbitrary
amplitude) and implemented as discussed in Sec. II-C.
The resulting time-integrated outgoing emission over a time well exceeding both the time
duration of the source and the time of propagation through the sample,
is displayed in Fig. 3 as a function of the number of subdivisions in each layer.
Specific modulations (called Kossel patterns) are observed at the Bragg angles of the multilayer.
These modulations are due to interferences of the diffracted waves inside the material \cite{Lan10,KLG19}. Outgoing
signals displayed in Fig. 3 are similar to the ones calculated for the same sample but by solving the
Helmholtz wave equation for a plane wave {\it incident} in the sample \cite{Pey20}, which is just a check of the
{\it optical reciprocity theorem} stated as {\it a point source at A will produce at B the
same effect as a point of equal intensity placed at B will produce at A} \cite{SB62,Born}.
Here, point A is a source point in the material while point B is a detection point at infinity.
Now, one can consider the opposite point of view of a plane wave originating from B and incident on the
material, and calculate the electric field in the material (supposedly a N-cell stack). The resulting
intensity (i.e. the sum of all of the local intensities) is equivalent to the total outgoing emission from
N identical sources located in the N cells of the material.

In Fig. 3, one sees a convergence in the number of
subdivisions necessary for performing accurate FDTD calculations of x-ray pulse propagation in multilayered materials.
It is interesting to note the extinction of Kossel structure $n=3$ for the converged results. This is consistent
with the diffraction theory when applied to a simple line grating \cite{Born} or to
X-UV interference mirrors \cite{Pardo88}.
Indeed, the first extinction should occur for the ratio $\frac{\Lambda}{e_{2}} = 3$. Taking for $e_{2}$ and
for the period $\Lambda=e_{1}+e_{2}$, the values given above, one finds a value very close to 3.
These different remarks validate our minimal implementation.

\begin{figure}
\rotatebox{0}{\resizebox{8.0cm}{6cm}{\includegraphics{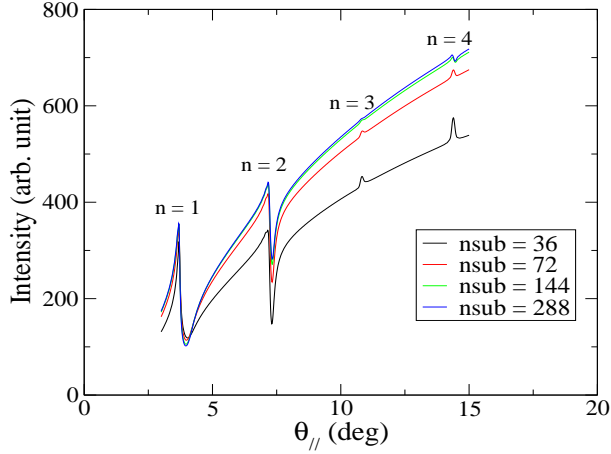}}}
\caption{\label{fig3} (Color online)
Calculated angular scan for the Mg
$K \alpha$ radiation emitted by a stack (Mg/Co)$_{30}$ ($e_{1}=5.45$ nm and $e_{2}=2.55$ nm)
where a same source of $K \alpha$ radiation has been put in each Mg layer.
Kossel patterns are labelled by their Bragg order $n$.
}
\end{figure}

\subsection{Propagation in an amplifying multilayer material}
We consider here the problem of a short pulse originating at the left, i.e. in the
vacuum cell of our computational domain (see Sec. II-B)
and then propagating from left to right
in a multilayer similar to the stack considered in the previous paragraph, albeit with
an increased number of bilayers, i.e. (Mg/Co)$_{60}$. The Mg layers are now {\it active}, i.e. described by
the Bloch equations, and we consider four typical sets of initial populations for $N_{1}$, $N_{2}$ in term
of the total density of atoms in solid state Mg that is n$_{Mg}$ = 4.3063 10$^{22}$ cm$^{-3}$. Population of
the lowest level is fixed to $N_{1}$ = 4.306 10$^{18}$ cm$^{-3}$ while $N_{2}$ is varied between
4.306 10$^{17}$ cm$^{-3}$ (case 1), 4.306 10$^{20}$ cm$^{-3}$ (case 2),
4.306 10$^{21}$ cm$^{-3}$ (case 3), 4.306 10$^{22}$ cm$^{-3}$ (case 4), respectively.
In this way, case 1 corresponds to a very weak 1s photoionization
with no inversion while case 4 corresponds to a maximal
population inversion in the Mg layers.
The ingoing pulse is of the form
$E(t)=\frac{1}{\sqrt{2\pi} \tau_{p}} \exp{[-0.5 (t-t_{o})^{2}}/\tau_{p}^{2}] \sin{\omega_{o}t}$
with the typical parameters $\tau_{p}=1$ fs and $t_{o}=2$ fs.
What is specifically studied here is the right outgoing intensity as a function of time.
More precisely, one displays the modulus of the Poynting vector, averaged over one period, i.e.
$S_{av}=\frac{1}{T} \int_{0}^{T} |S| dt$ with $\mu_{o}^2S^{2}=(E_{x}B_{z})^{2}+(E_{x}B_{y})^{2}$,
$B_{z}=\frac{1}{c}E_{x} \sin{\theta_{\bot}}$. Hereafter, units for the Poynting vector are
the atomic units, i.e. $S_{av}$ is in unit of
$\frac{e^{12}}{(4\pi \epsilon_{o})^6} \frac{m_{e}^{4}}{\hbar^9}$.
For a signal ingoing in the normal direction,
calculations are diplayed in Fig. 4.

\begin{figure}
\rotatebox{0}{\resizebox{8.0cm}{6cm}{\includegraphics{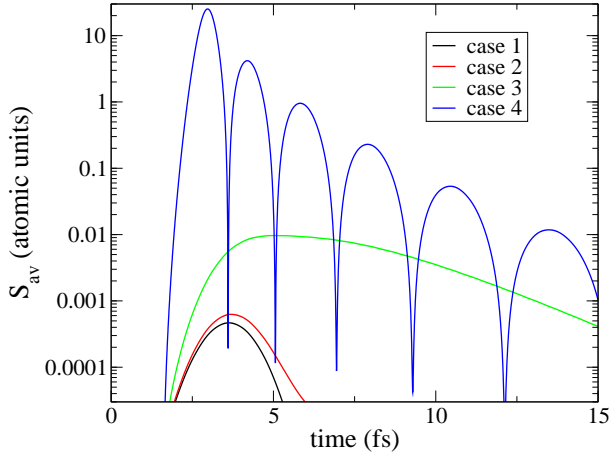}}}
\caption{\label{fig4} (Color online)
Modulus of the Poynting vector outgoing from the stack (Mg/Co)$_{60}$ in the
normal direction ($\theta_{//}=90^{o}$),
as a function of time.
See text for the characterics of the ingoing signal and for the definition of cases 1-4.
}\vspace{0.5cm}
\end{figure}
Compared with the weak signal (case 1), one sees the gradual effect of a gain material on the
intensity temporal shape of the outgoing pulse. In particular, one notices an increase of the
outgoing pulse duration with respect to the ingoing pulse duration.
In the case of strong (and here
maximal) inversion (case 4), a typical effect such as "ringing" of the outgoing signal is observed.
This correspond to the well-known Burnham-Chiao ringing \cite{BC69}. Present simulations are in
the time domain. Of course, taking the Fourier transform gives information on the frequency domain.
Fig. 5 shows the spectra corresponding to the previous simulations, i.e. the evolution of the
(normalized) spectrum as a function of the density of inversion.
Even on this small distance of propagation, the
high density allows a clear gain narrowing (case 2 and case 3)
and then a strong AC Stark (or Rabi) splitting for the
maximal density of inversion (case 4). This behavior illustrates the response of the two-level system
driven by an x-ray field on resonance. This field becomes so important that the levels shift
dynamically through the Stark effect.

\begin{figure}
\rotatebox{0}{\resizebox{8.0cm}{6cm}{\includegraphics{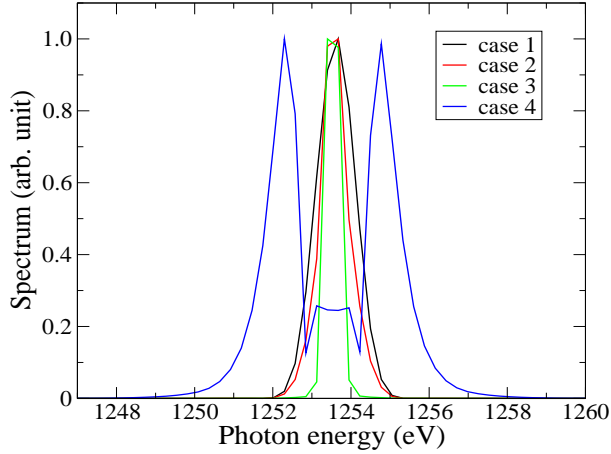}}}
\caption{\label{fig5} (Color online)
Normalized spectra corresponding to the signals of Figure 4.
}\vspace{0.5cm}
\end{figure}

\subsection{Self-emission of an amplifying multilayer material}
In this paragraph, we do not consider the propagation of an external pulse but the signal originating
from the noisy source of spontaneous emission in each active cell of a multilayer.
More precisely, we study how spontanenous emission emitted in one direction propagates and how
stimulated emission sets in.
Present calculations rely on the modeling of spontaneous emission presented in Sec. II-C.
Still for the
same multilayer (Mg/Co)$_{60}$ and the same sets of initial populations ($N_{1}$, $N_{2}$) in
Mg layers,
Fig. 6 displays simulations of the outgoing X-ray emission in the
normal direction.

\begin{figure}
\rotatebox{0}{\resizebox{8.0cm}{6cm}{\includegraphics{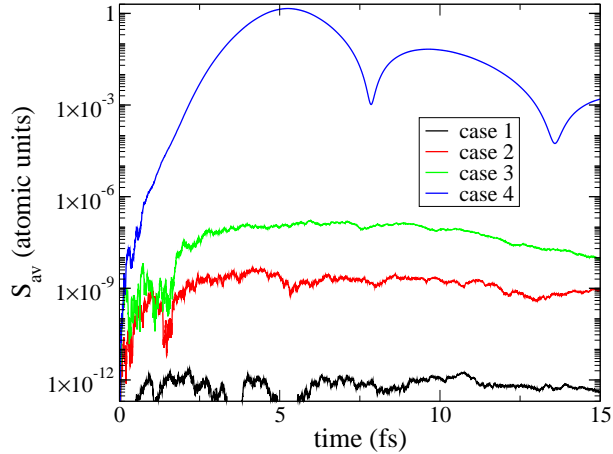}}}
\caption{\label{fig6} (Color online)
Modulus of the Poynting vector outgoing from the stack (Mg/Co)$_{60}$ in the normal
direction ($\theta_{//}=90^{o}$),
as a function of time. The seeding is provided by the inherent spontaneous emission noise.
See text for the for the definition of cases 1-4.
}
\end{figure}
\noindent We see the weak noisy signal for a low initial excitation (case 1) while, gradually with the
density of inversion, a collective emission is set up independently of any external pulse ingoing
into the material.
For the maximal initial population inversion (case 4), characteristics of the superradiance,
namely time delay for the peak of emission and ringing are clearly visible.
We emphazise that these results are just single realizations which may be subjected to large fluctuations.
However, what we observed is that in case of strong population inversion, stimulated emission and
amplification make the results weakly sensitive to a given realization. A fact which is somehow
reflected by the smooth aspect of graph corresponding to case 4. In case of weak or absent stimulated
emission, the right temporal behavior can be recovered only by performing an average over many
realizations. For case 1, 2, we checked that an average over at least a few tens of realizations gives
a decaying behavior for large time. Furthermore, in this weak limit, we observed a behavior
as $t e^{-\Gamma t}$ ($\Gamma$ being the decay time of level 2) instead of the right behavior
$e^{-\Gamma t}$ which is an identified shortcoming of MB models with a noise term \cite{Kru18,Ben19}.
In the weak limit, present theoretical approach is thus not very reliable.

\noindent We turn now (Fig. 7) to a propagation seeded around the first uncorrected Bragg angle
(defined so that $\Lambda sin{\theta_{//}} = \lambda/2$, $\Lambda$ being the period of the
material, i.e. $\Lambda = e_{1}+e_{2}$ according to Fig. 2).
Here
an oscillation feedback can be provided by Bragg reflection \cite{Yariv74, Yariv77, JMA14} so that
large electric field enhancements can be obtained \cite{Pey20}.
Note that, for a multilayer and in the linear response regime, a small deviation to the
previous Bragg law exists \cite{Atwood, Chant95, Pardo88}.
Compared with Fig. 6 (normal direction), a dramatic change of the emission
is observed for $\theta_{//}= 3.45^{o}$.
Here at oblique incidence, the
phase front of a plane wave incident on the left is supposed to arrive on
the opposite side in a very short time.
This time $\tau$ is defined so that
$c\tau = d \sin{\theta_{//}}$, $d$ being the whole thickness of the multilayer, i.e.
$\tau \simeq 0.1 fs$ in present conditions.
As one can see, depending on the inversion
density, the outgoing x-ray pulse shifts ealier in time, its duration is reduced while its intensity strongly
increases. This is a clear evidence of the feedback provided by Bragg reflection in the multilayer. Some
complex "ringing" is also apparent.
Here also, we observed that stimulated emission, amplification and feedback make the results weakly sensitive
to a particular realization which means that the random onset of emission is easily forgotten.
Spectra corresponding to Fig. 7 are displayed in Fig. 8. The huge broadening observed
for the maximal density of inversion (case 4) is a combined effect of the pulse shortening and
of AC Stark splitting due to the huge electric field which sets up.
These results suggest that for large population inversions, emission will tend to be around the Bragg angle and that
its duration should be extremely short, beating even Auger decay which is of the order of 2-3 fs.

\begin{figure}
\rotatebox{0}{\resizebox{8.0cm}{6cm}{\includegraphics{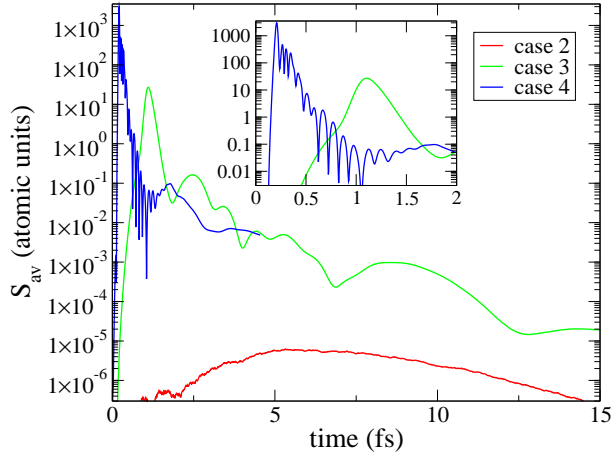}}}
\caption{\label{fig7} (Color online)
Modulus of the Poynting vector outgoing from the stack (Mg/Co)$_{60}$ in first Bragg
direction $\theta_{//} = 3.45 \deg$,
as a function of time. The seeding is provided by the inherent spontaneous emission noise.
See text for the for the definition of cases 2-4. The inset is a zoom of the time interval 0-2 fs.
}\vspace{0.7cm}
\end{figure}

\begin{figure}
\rotatebox{0}{\resizebox{8.0cm}{6cm}{\includegraphics{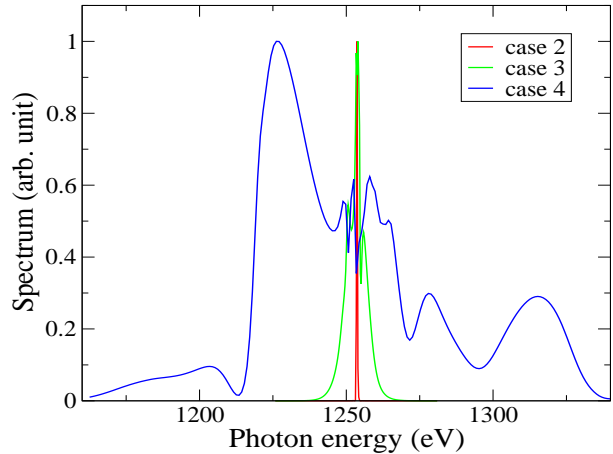}}}
\caption{\label{fig8} (Color online)
Normalized spectra corresponding to the signals of Figure 7.
}\vspace{0.5cm}
\end{figure}

At this step, it is important to comment the choice of the relative thicknesses which can be done to
optimize the feedback.
For our purpose, optimization of the multilayer relies on its intrinsic reflectivity at the wavelength of interest
(i.e. here the $K \alpha$ line of active layers) regardless any MB calculation. Relative thicknesses
are thus important. A
characteristic parameter is the ratio (usually
called $\gamma$) of the heaviest layer (and more absorbing) thickness $e_2$ to the
period. It is well known from multilayer physics that an optimal diffraction may be found for $\gamma$ close to 1/3.
Using equations based on two different methods (either the dynamical theory of diffraction \cite{Batt64}
as applied to multilayer optics \cite{Koz87}) or the "optical approach" as applied to multilayer optics
\cite{Pardo88}, we have found that, for the system (Mg/Co) the optimal $\gamma$ is close to 0.4. So the
rule of thumb of $\gamma \sim 1/3$ is verified and has been used as a typical value.
Concerning the effect of possible imperfections in the periodicity, one can also look at the intrinsic reflectivity
and see how it is sensitive to a small random variation of thicknesses. Considering 1$\%$ variation
(which is a high limit from an experimental point of view)
we have found that the reflectivity moves of only 1$\%$. Therefore, small random variations of the thicknesses
are not very important.


\begin{figure}
\rotatebox{0}{\resizebox{8.0cm}{6cm}{\includegraphics{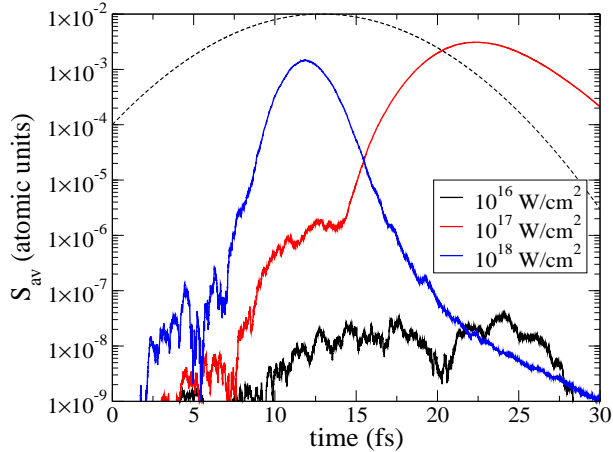}}}
\caption{\label{fig9} (Color online)
Modulus of the Poynting vector outgoing from the stack (Mg/Co)$_{60}$ in the
normal direction ($\theta_{//}=90^{o}$),
as a function of time and for different intensities of an external X-ray pump. Dashes (of arbitrary
unit) indicate
the temporal shape of the pump.
The seeding of the emission is provided by the inherent spontaneous emission noise.
}
\end{figure}

\begin{figure}
\rotatebox{0}{\resizebox{8.0cm}{6cm}{\includegraphics{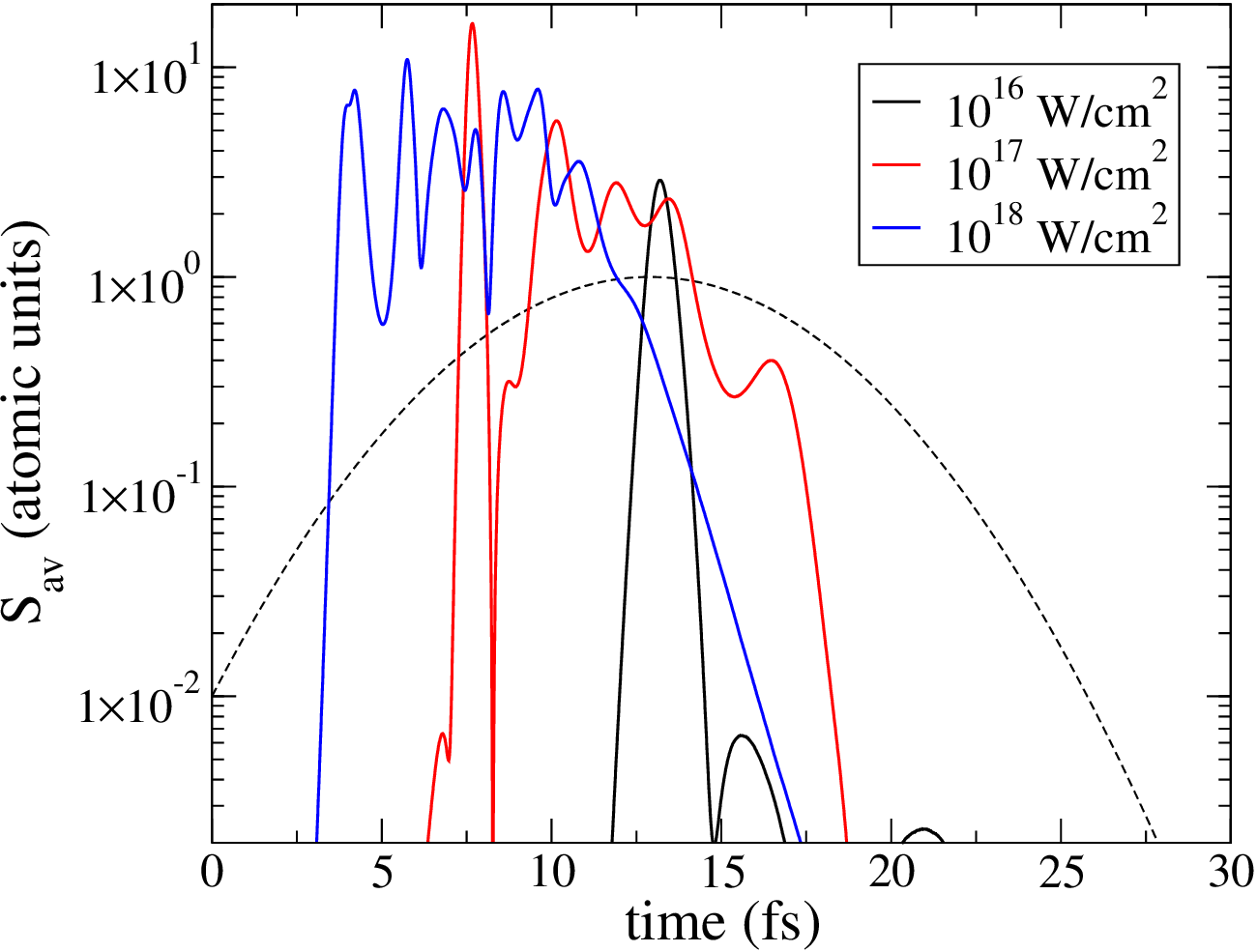}}}
\caption{\label{fig10} (Color online)
Modulus of the Poynting vector outgoing from the stack (Mg/Co)$_{60}$ in the
direction $\theta_{//} = 3.45^{o}$,
as a function of time and for different intensities of the external X-ray pump. Dashes (of arbitrary
unit) indicate the temporal shape of the pump.
The seeding of the emisssion is provided by the inherent spontaneous emission noise.
}\vspace{0.5cm}
\end{figure}

\subsection{Self-emission of a pumped multilayer material}
Previous calculations were based on the idea of
 a preliminary preparation of $N_{1}$, $N_{2}$ at given initial values.
Here we place ourselves in conditions where these initial values are zero and where pumping of level 2
is provided by an external x-ray source photoionizing the ground level (of population $N_{0}$)
within an active layer, as discussed at the beginning of Sec. III.
This rises the question of the optimal size of the multilayer, i.e. of the number of bilayers.
The answer depends on the material, on the attenuation length of the pump and on the incidence angle.
For a not too inhomogeneous pumping, the size of the multilayer for a given incidence angle cannot
exceed the attenuation length of the pump.

In present
simulations, the x-ray pulse is supposed Gaussian and of 10 fs duration (FWHM) and at normal incidence,
i.e. according to Fig. 2. Its intensity is propagated and depleted according to Eq. (20).
The photon energy of this
pump is $1332$ eV, i.e. above the Mg K-edge. In Fig. 9, we plot the outgoing Mg $K \alpha$ emission
of our (Mg/Co)$_{60}$ stack in
the normal direction, as resulting from three different pump intensities.
Dashes (of arbitrary unit)
indicate the temporal shape of the pump. Here, the different shapes of the outgoing
emission reflect the efficiency of the core-hole creation with respect to the maximum of intensity. As seen
above, the signal stops to be noisy when stimulated emission sets up which is possible if a
sufficient density of core-holes is reached. Now, as expected from the considerations of Sec. III-C, the
signals observed in the Bragg direction $\theta_{//} = 3.45^{o}$, are dramatically different (Fig. 10) both
in intensity (few orders of magnitudes) and in temporal shape.
About the different angles of emission, a question arise here.
In principle, spontaneous emission (which is isotropic) generates photons in all directions although at
different times and different locations.
For this reason, the fields corresponding to all directions should be calculated so that
Eqs 4-6 include the contribution of all directions.
In term of computation time, such a treatment is prohibitive since, as discussed in Sec. III-A and seen in Fig. 3,
a very fine spatial zoning and a very fine angular gridding is needed to resolve
the Kossel structures.
Here our purpose is to show that a multilayer structure may present preferential
directions of emission for which initial photons emitted around the Bragg angle "catch on" the stimulated
emission as does a standard cavity. This point is illustrated in Fig. 11 which, for the same multilayer and
for the intensity $10^{17}$ W/cm$^{2}$, displays
at $\theta_{//} = 45^{o}$ and for 3 angles
at and around the Bragg angle, the temporal profiles of radiation. What is shown in this figure in twofold.
First, one sees that most of the emission is catch on around the Bragg angle $\theta_{//} = 3.45^{o}$.
Second, it is clear that the enhancement of the emission is not a simple geometry effect where this enhancement
would come from a larger effective propagation distance.\\

In Fig. 10, the complex behavior of outgoing signals
reflects the complex interplay between population kinetics, depletion of the pump, propagation and Bragg
diffraction. This is somehow illustrated by
snapshots of the spatial distribution (inside the multilayer) of
populations $N_{0}$, $N_{1}$, $N_{2}$ at different characteristic times during the driving pump pulse
(Fig. 12). The x-ray pump comes on the
right so that the decrease of population $N_{0}$ as a function of $z$ reflects the attenuation of
the pump as it propagates from the right to the left inside the material.
The Maxwell-Bloch coupling of populations
 $N_{2}$, $N_{1}$, seeded by spontaneous emission, gives rise to a set of pulses propagating
inside the multilayer and to clear Rabi oscillations (Rabi flopping).
Despite the complex temporal shape of these
outgoing signals around the Bragg direction, they remain much more intense than in the normal direction.

\begin{figure}
\rotatebox{0}{\resizebox{8.0cm}{6cm}{\includegraphics{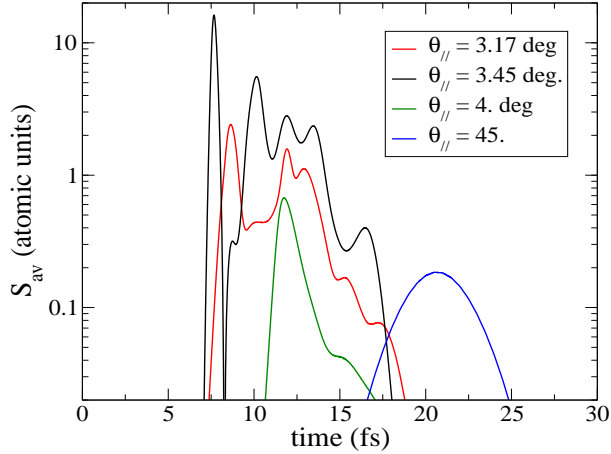}}}
\caption{\label{fig11} (Color online)
Modulus of the Poynting vector outgoing from the stack (Mg/Co)$_{60}$ in
four directions,
as a function of time and for the intensity $10^{17}$ W/cm$^{2}$ (of external X-ray pump).
The seeding of the emisssion is provided by the inherent spontaneous emission noise.
}\vspace{0.5cm}
\end{figure}

\begin{figure}
\rotatebox{0}{\resizebox{12.0cm}{6cm}{\includegraphics{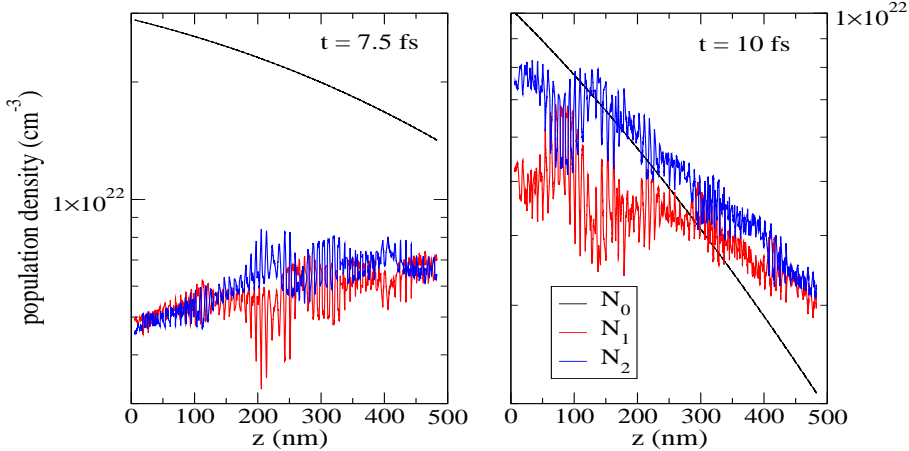}}}
\caption{\label{fig12} (Color online)
Snapshots of populations $N_{0}$, $N_{1}$, $N_{2}$ at 2 moments inside the
multilayer (Mg/Co)$_{60}$ and in the conditions of Fig. 10. The pump comes on the
right with an intensity of $10^{18}$ W/cm$^{2}$.
Left figure: $t=7.5$ fs; right figure $t=10$ fs.
}
\end{figure}
As massive photoionization causes a large concentration of photoelectrons which thermalize quickly in dense
materials, an interesting question is, to what extent collisional ionization of 2p electrons in level 2
by these photoelectrons may affect significantly the lifetime of this level.
Assuming an instantaneous thermalization, a temperature for these photoelectrons can be estimated from
the total energy deposited in these photoelectrons.
In Mg and with our parameters, we estimated this electron temperature to be of the order of 100 eV
for a pump intensity of $10^{18}$ W/cm$^{2}$.
From a calculation of the collisional ionization cross section of a 2p electron in state
$1s\ 2s^{2}\ 2p^{6}\ [3s^{2}]$,
the collisional ionization rate varies between $10^{9}$ s$^{-1}$ (at 10 eV) to
$10^{14}$ s$^{-1}$ (at 100 eV). This can be compared with the autoionization rate 
which is $3.4 \ 10^{14}$ s$^{-1}$. Although not negligible at high intensity, and still,
at the end of pump pulse, the effect of photoelectrons remains smaller than autoionization decay.
We do not think that their existence changes the conclusions of this article especially in the
feedback regime which shortens the emission duration.

\subsection{Self-emission of a pumped natural crystal}
In this last section, one examines the case of a natural crystal whose periodicity of atomic layers
may provide the same kind of Bragg oscillations. One considers here a Ni crystal where for an
orientation (111) of planes parallel to the surface, atomic layer spacing is $d=0.216$ nm.
A strong pumping of 1s core electrons in Ni may give rise to an amplification on the $2p \rightarrow 1s$
$K \alpha1$ line at 7478.15 eV. At this energy the first Bragg angle is around 22.6$^{o}$ (with respect to
the surface).
1D Periodicity is introduced in the calculations by considering the (supposedly perfect) crystal
as a stack of bilayers of period $d$ and where the first layer (the active layer) is a layer of Ni atoms
while the second layer is just empty (then passive) and of refractive index 1.
Such a replacement of real atoms by a uniform layer of a given thickness $e_1$ (See Fig. 2)
is a rough method to simulate
the problem of a distribution of individual small scatterers. Its validity is semi-empirical.
A relevant quantity to measure its effectiveness is the reflectivity of the system (element/vacuum)$_n$ which
for a wavelength of interest can be calculated around the Bragg angle and compared with a model giving the
reflectivity of a real crystal.
From x-ray reflectivity calculations based on a solution of the Helmholtz equation applied to stacks
(element/vacuum)$_n$, we performed such comparisons with a software available at Sergey Stepanov's X-ray server
\cite{Stepanov} allowing a calculation of the reflectivity in real crystals.
We found that, taking for the element thickness a typical value of 0.4 $d$ ($d$ being the proper inter-reticular
distance in the crystal of interest) and renormalizing properly the number of atoms in this element layer
to the right number of atoms (per volume unit), this approach gives results close to those of the
Stepanov's X-ray server. Accordingly, we used this recipe in our Maxwell-Bloch calculations.

Simulations presented here correspond to
a Ni thickness of 1.08 $\mu m$, i.e. to the stack (Ni/vacuum)$_{5000}$.
Simulation results are displayed in Fig. 13. Irradiation conditions are a raised cosine pulse of
10 fs duration (FWHM), 8360 eV of photon energy (i.e. above the Ni K-edge) and of intensities $10^{19}$
W/cm$^{2}$, and at normal incidence, i.e. in the geometry of Fig. 2.
Fig. 13 displays a comparison of the outgoing Ni $K \alpha1$
signal (i.e. seeded by spontaneous emission)
observed in the normal direction $\theta_{//}=90^{o}$, in the Bragg
direction $\theta_{//}=21.57^{o}$ and in some off-Bragg direction,
respectively. Compared with the other directions, emission in the Bragg diffraction region (the Kossel
region) is strongly enhanced which indicates the possibility of having
a resonator or feedback effect in a natural crystals.

\begin{figure}
\rotatebox{0}{\resizebox{8.0cm}{6cm}{\includegraphics{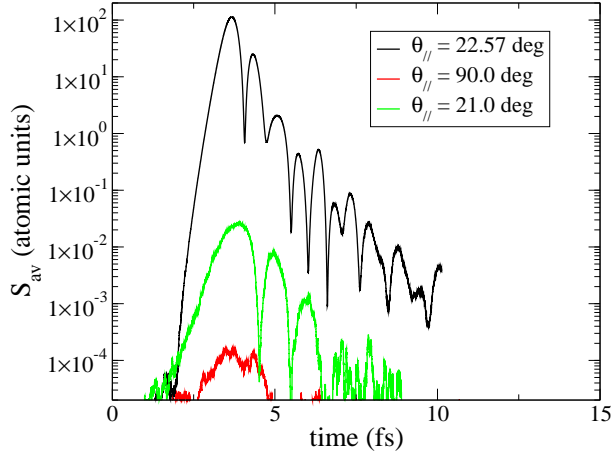}}}
\caption{\label{fig13} (Color online)
Modulus of the Poynting vector outgoing from the stack (Ni/vacuum)$_{5000}$
($e_{1}=0.0972$ nm, $e_{2}=0.1188$ nm) irradiated in the normal direction
by a raised cosine pulse of
intensity $10^{19}$ W/cm$^{2}$, 10 fs duration, 8360 eV photon energy.
Three outgoing directions are shown: the normal direction $\theta_{//} = 90^{o}$, the
Bragg direction $\theta_{//} = 22.57^{o}$ and an off-Bragg axis direction
$\theta_{//} = 21.0^{o}$.
Seeding of the emisssion is provided by the inherent spontaneous emission noise.
}
\end{figure}

\section{conclusion}

A 1D Maxwell-Bloch FDTD model for any oblique incidence has been successfully implemented for studying
x-ray propagation in 1D photonic crystals.
We simulated the self-emitted signal from typical 1D photonic crystals where a population inversion is
prepared on an atomic transition in the x-ray range. The build up of outgoing signal
starts from spontaneous emission.
We have seen that this emission encompasses many non-linear phenomena
such as Rabi splitting, Rabi flopping, ringing, etc in the x-ray range as well as the
Kossel effect but in an {\it amplified mode}.
We have shown that most of the emission occurs in a prevailing direction
which is the Bragg direction. If the inversion results from a previous photoionization, we observed
that this emission is short enough to beat Auger relaxation.
We specifically studied cases where the pumping source allowing a strong population inversion
is an intense, short x-ray pulse such as provided by XFEL sources. For a typical multilayer and for
 realistic conditions of pumping, calculations show a strong enhancement of the emission in the
Bragg direction. For the case of natural crystals, this enhancement is also noticeable.

Results of this study motivate future experimental investigations of the behavior of photonic crystals
whether they are natural or artificial (multilayers). It motivates also many other
theorerical investigations on
different multilayers or natural crystals to optimize x-ray emission at different wavelengths.

\acknowledgements
At numerous times during the course of this work, one of us (O. Peyrusse)
has benefited from discussions and
helpful advices concerning the use of computational resources from Paul Genesio at
PIIM laboratory.


\newpage 


%

\end{document}